\documentclass[prl,twocolumn,superscriptaddress,floatfix,graphicx,showpacs]{revtex4-2}
\usepackage{ifpdf}
 \newif\ifpdf
\ifx\pdfoutput\undefined
   \pdffalse
\else
   \pdfoutput=1
   \pdftrue
\fi
\ifpdf
   \usepackage{graphicx}
   \usepackage{epstopdf}
   \usepackage{color}
\else
   \usepackage{graphicx}
\fi

\usepackage{amsmath,amssymb}

\usepackage{epsfig}

\begin{document}

\title{Freezing of two-length-scale systems: complexity, universality and prediction }

%\title{The uncertainty of glass transition temperature $T_g$ in molecular dynamics simulations and numerical algorithm for unique determining of $T_g$}

\author{R.E. Ryltsev}
\affiliation{Institute of Metallurgy, Ural Branch of Russian Academy of Sciences, 620016, 101 Amundsena str., Ekaterinburg, Russia}
\affiliation{Ural Federal University, 620002, 19 Mira str., Ekaterinburg, Russia}
\affiliation{Institute for High Pressure Physics, Russian Academy of Sciences, 142190 Troitsk, Russia}

\author{N.M. Chtchelkatchev}
\affiliation{Institute for High Pressure Physics, Russian Academy of Sciences, 142190 Troitsk, Russia}

\author{V. Ankudinov}
\affiliation{Institute for High Pressure Physics, Russian Academy of Sciences, 142190 Troitsk, Russia}

\author{V.N. Ryzhov}
\affiliation{Institute for High Pressure Physics, Russian Academy of Sciences, 142190 Troitsk, Russia}

\author{M. Apel}
\affiliation{Access e.V. – Materials and Processes
An-Institut der RWTH Aachen, 52074 Aachen, Germany}

\author{P.K. Galenko}
\affiliation{Faculty of Physics and Astronomy, Otto Schott Institute of Materials Research, Friedrich-Schiller-Universität-Jena, 07743 Jena, Germany}

\begin{abstract}
Two-length-scale pair potentials arise ubiquitously in condensed matter theory as effective interparticle interactions in molecular, metallic and soft matter systems. The existence of two different bond lengths generated by the shape of potential causes complex behavior in even one-component systems: polymorphism in solid and liquid states, water-like anomalies, the formation of quasicrystals and high stability against crystallization. Here we address general properties of freezing in one-component two-length-scale systems and argue that the formation of solid phases during cooling a liquid is essentially determined by the radial distribution function (RDF) of the liquid. We show that different two-length-scale systems having similar RDF freeze into the same solid phases. In most cases, the similarity between RDFs can be expressed by the proximity of two dimensionless effective parameters: the ratio between effective bond lengths, $\lambda$, and the fraction of short-bonded particles $\phi$. We validate this idea by studying the formation of different solid phases in different two-length-scale systems. The method proposed allows predicting effectively the formation of solid phases in both numerical simulations and self-assembling experiments in soft matter systems with tunable interactions.
\end{abstract}

\maketitle
\section{Introduction}

 The behavior of even very complex condensed matter systems can be qualitatively (and sometimes quantitatively) described using coarse-grained effective pair potentials. Such potentials should be considered as a result of coarsening the real interaction by averaging over a certain set of variables, which are excluded from consideration \cite{Likos2001PhysRep}.

For example, by averaging over angular variables of polar molecules with non-isotropic interaction, it is possible to obtain an isotropic effective pair potential and consider the molecular system as a set of ''atoms'' interacting through such a potential~\cite{Mishima1998Nature}. Another example is the effective pairwise interaction potentials between colloidal particles~\cite{Likos2001PhysRep}. In reality, colloidal suspensions are complex multicomponent systems including different time-space scales~\cite{morrison2002colloidal}. For example, a solution of sterically stabilized colloidal particles is a system consisting of a molecular solvent (for example, water), nano- or micro-sized colloidal particles, as well as polymer molecules that are introduced onto the surface of the colloidal particles to prevent coagulation. If we are interested in, for example, crystallization processes of colloidal nanoparticles or short-range order in a fluid state, then a colloidal suspension can be considered as a system of ''atoms'' interacting through some effective potential that describes the interaction between colloidal particles at spatial scales comparable to their sizes.

%(see Fig. ~ \ ref {fig: water_2scale_pot}, Fig. ~ \ ref {fig: soft_matt_pot})

An analysis of the available effective pair potentials allows concluding that very many such potentials have more than one length scale, that is, they generate several characteristic interparticle distances determined by the presence of several attractive wells, negative curvature in the repulsion region, nonmonotonicity of the derivatives, or due to the ultrasoftness of the repulsive part. That is true for molecular systems~\cite{Jagla1998PRE,Oliveira2008JCP,Mishima1998Nature,Yan2005PRL,Yan2008PRE,Gribova2009PRE}, metals and their alloys~\cite{Lee1981Book,McMahan1983PRB,Dzugutov1991JNCS,Mihalkovic20012PRB,Englel2015Nature} and soft matter systems~\cite{Likos1998PRL,Likos2002JPCM, Watzlawek1999PRL,Mayer2007Macromol,Rechtsman2006PRE,Louis2002PRE,Likos2001PhysRep,Komarov2018SoftMatt}.  Thus, multiscale effective interactions is a common property of systems of various nature.

The existence of two different bond lengths generated by the shape of potential causes complex phenomena in even one-component systems, such as polymorphism in solid \cite{Fomin2008JCP,Englel2015Nature}, liquid \cite{Xu2006JPhysCondMatt,Xu2009JChemPhys} and glassy~\cite{Xu2011JChemPhys,Buldyrev2009JPhysCondMatt} states, water-like anomalies\cite{Oliveira2008JCP,Mishima1998Nature,Yan2005PRL,Yan2008PRE,Gribova2009PRE,Vilaseca2011JNonCrystSol,Kumar2005PRE}, the formation of quasicrystals \cite{Dzugutov1993PRL,Engel2007PRL,Archer2013PRL,Barkan2011PRB,Dotera2014Nature,Engel2014PRL,Ryltsev2015SoftMatt,Ryltsev2017SoftMatt,Englel2015Nature,Damasceno2017JPCM,Kryuchkov2018SoftMatt} and high stability against crystallization~\cite{Ryltsev2013PRL}.

An important issue is if there are any universal properties of multi-length-scale systems, which do not depend on a particular form of the potential. Another point is how to predict the formation of solid phases by analyzing the properties of the liquid. Recently, we addressed these issues on an example of quasicrystals formation in two-length-scale systems~\cite{Ryltsev2015SoftMatt,Ryltsev2017SoftMatt}. We found that the formation of decagonal and dodecagonal quasicrystals is universally determined by the values of two dimensionless effective parameters: the ratio between effective bond lengths, $\lambda$, and the fraction of short-bonded particles $\phi$. Here we show that this idea can be successfully applied for studying crystallization

\section{Methods}

We investigate by the molecular dynamics simulations one-component 3D systems of particles interacting through different two-length-scale potentials. These include the well-known Dzugutov potential~\cite{Dzugutov1992PRA}, repulsive shoulder system (RSS) potential~\cite{Fomin2008JCP,Gribova2009PRE}, modified oscillating pair potential (OPPm)~\cite{Ryltsev2015SoftMatt,Ryltsev2017SoftMatt} (which is slightly modified potential which was first introduced in \cite{Mihalkovich2012PRB} and then used to simulate icosahedral QCs~\cite{Englel2015Nature}), Yoshida-Kamakura potential~\cite{Yoshida1976ProgTheorPhys}.

Hereafter we use dimensionless units like Lennard-Jones ones that is normalizing the energy, temperature and distance by the corresponding potential parameters. For example, for RSS we have $\tilde{{\bf r}}\equiv {\bf r}/d$, $\tilde U=U/\varepsilon$, temperature $\tilde T=T/\epsilon$, density $\tilde{\rho}\equiv N d^{3}/V$, and time $\tilde t=t/[d\sqrt{m/\varepsilon}]$, where $m$ and $V$ is the molecular mass and system volume correspondingly. For the EAM model of aluminum, the value of effective pair potential $U_{\rm eff}$ at the first minimum was chosen as the energy unit.

For molecular dynamics simulations, we use $\rm{LAMMPS}$ package~\cite{Plimpton1995JCompPhys}. The system of $N=20000$ particles was simulated under periodic boundary conditions Nose-Hoover NVT ensemble.   This amount of particles is enough to obtain satisfactory diffraction patterns to study (quasi)crystal symmetry~\cite{Englel2015Nature,Ryltsev2015SoftMatt,Ryltsev2017SoftMatt}. The molecular dynamics time step was $\delta t=0.003-0.01$ depending on system temperature \cite{Kuksin2005MolSim, Norman2001JETP}.

To study solid phases, we cooled the system starting from a fluid in a stepwise manner and completely equilibrated at each step. The time dependencies of temperature, pressure and configurational energy were analyzed to control equilibration \cite{Kuksin2005MolSim}.

%\textcolor{red}{few words about choose of time-scales, control undercooling by MSD,...}

To study the structure of both fluid and solid phases we use radial distribution functions $g(r)$, bond order parameters $q_l$~\cite{Steinhardt1981PRL,Steinhardt1983PRB,Hirata2013Science}, diffraction analysis and visual analysis of the snapshots. Detailed description of these methods as well as the procedure for preparing and relaxing the solid phases are presented in Ref.~\cite{Ryltsev2015SoftMatt}.

\section{Pair correlation functions and effective parameters\label{sec:eff_par}}

 \begin{figure}
  \centering
  \includegraphics[width=\columnwidth]{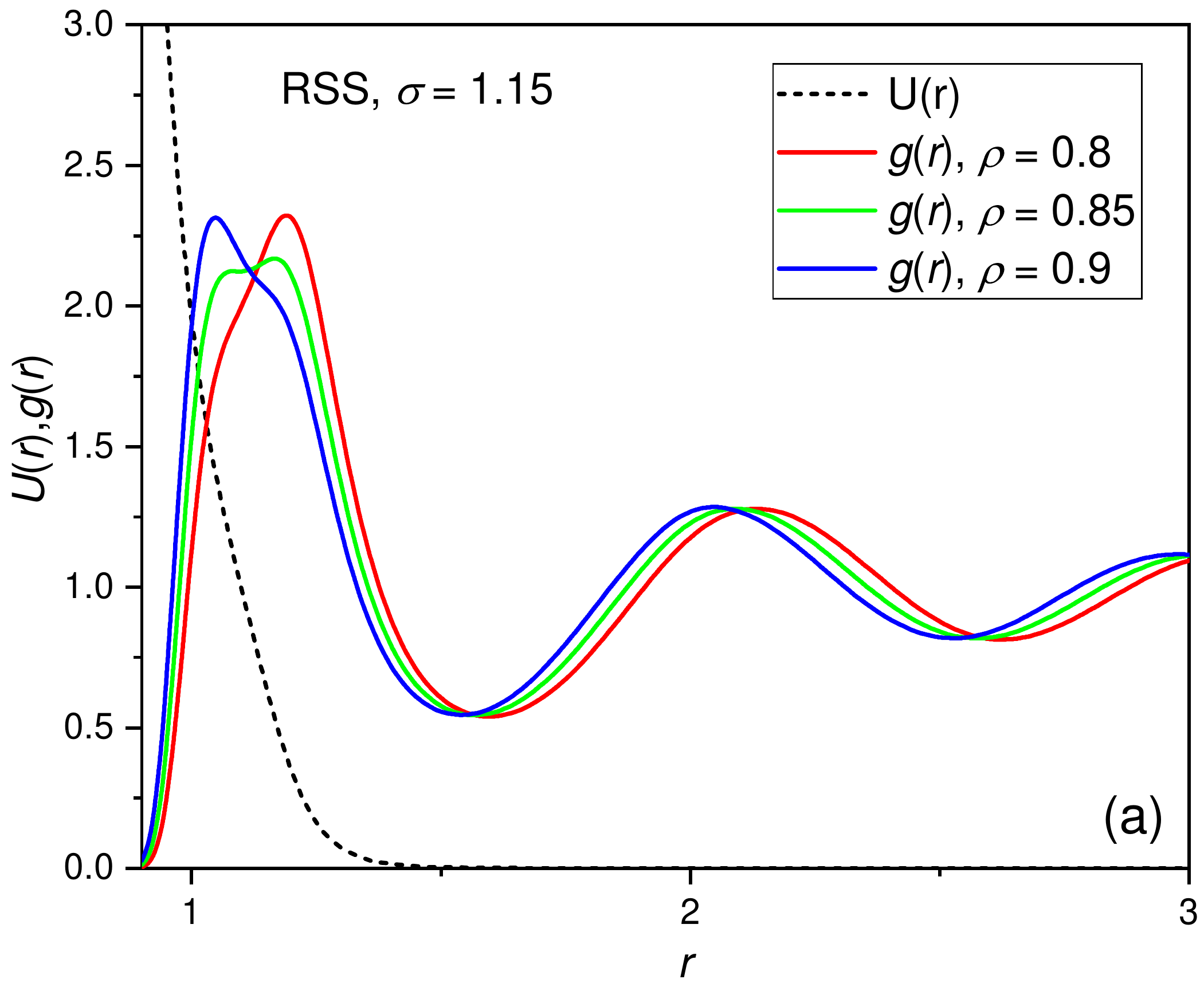}\\ \includegraphics[width=\columnwidth]{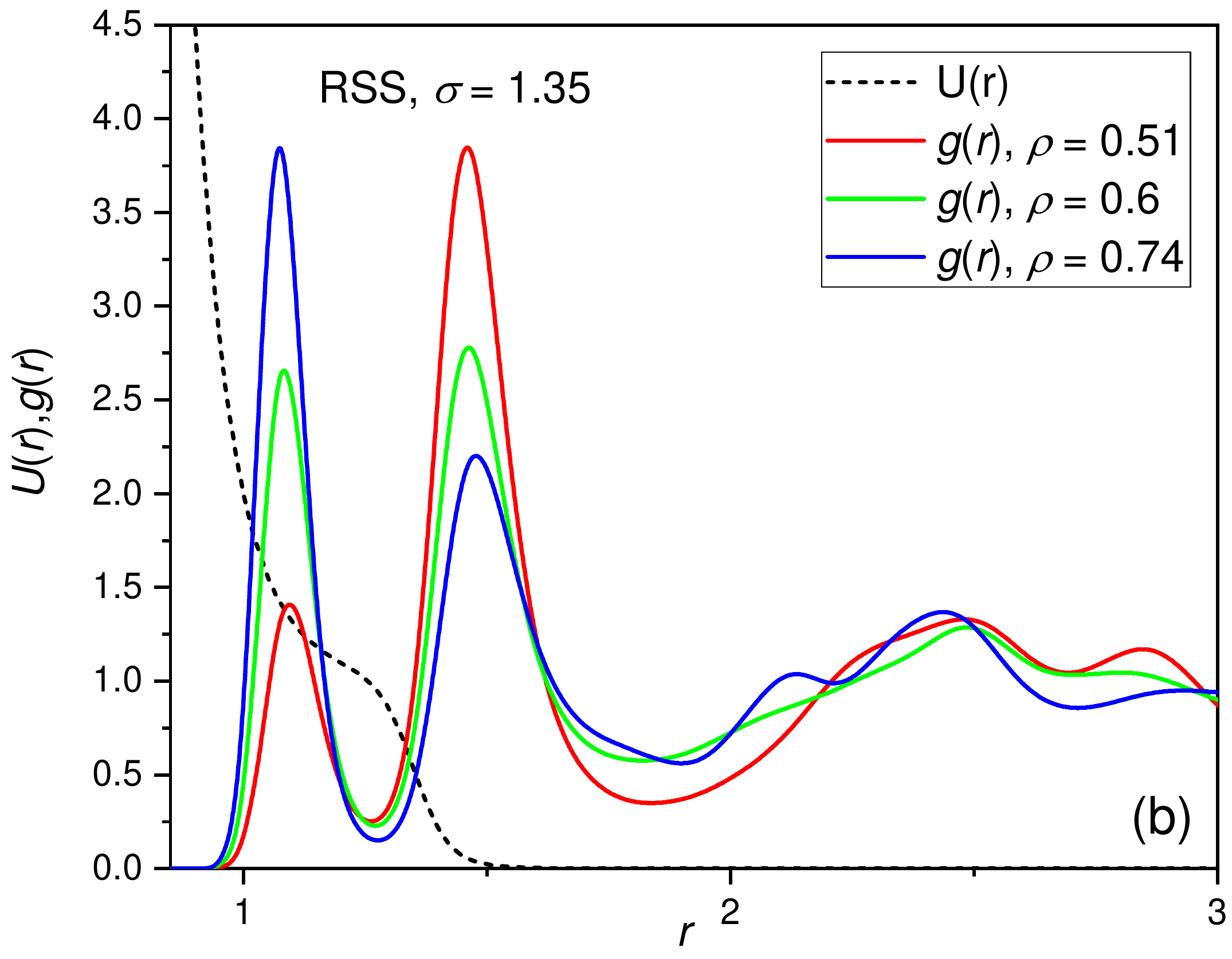}\\ \includegraphics[width=\columnwidth]{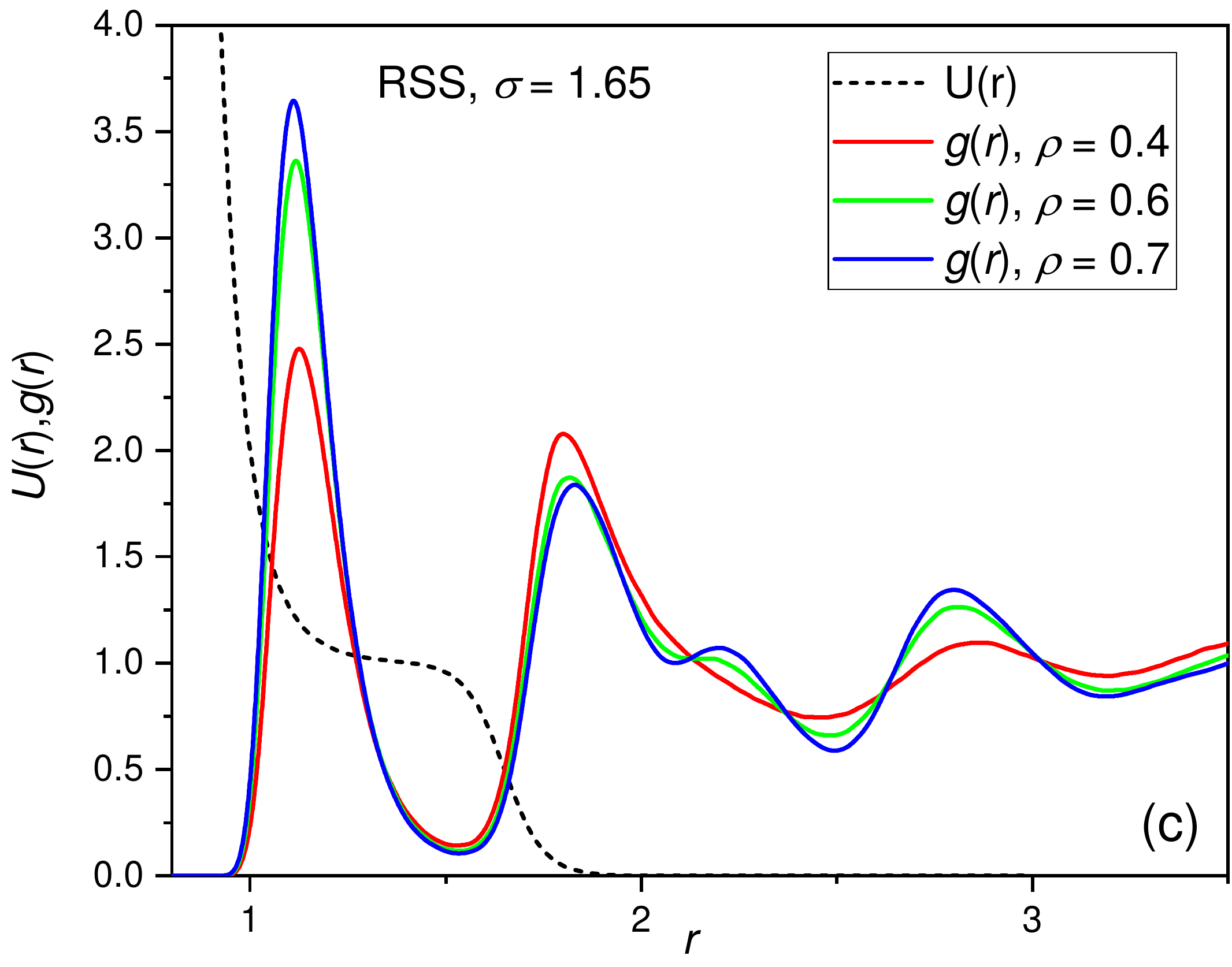}\\
 \caption{Radial distribution functions and potential curves for RSS model at different widths of repulsive shoulder $\sigma$: (a) $\sigma=1.15$; (b) $\sigma=1.35$, $\sigma=1.35$.}
  \label{fig:eff_param}
\end{figure}

The main idea we suggest is that the symmetry of solid phases which form at freezing is essentially determined by radial distribution function (RDF) of high-temperature (liquid or supercooled liquid) phase. In other words, systems having similar RDFs in liquid phase freeze into the similar solid phases~\cite{Ryltsev2015SoftMatt,Ryltsev2017SoftMatt}.

In the case of two-length-scale systems, the similarity between RDFs can be expressed by the proximity of two dimensionless effective parameters: the ratio between effective bond lengths, $\lambda$, and the fraction of short-bonded particles $\phi$~\cite{Ryltsev2015SoftMatt,Ryltsev2017SoftMatt}. These parameters can be extracted from liquid phase RDFs. Indeed, RDF peaks in two-length-scale systems usually demonstrate special  splitting of RDF peaks or existence of shoulders which express the existence of two characteristic bond lengths. Thus, the ratio between effective bond lengths, $\lambda$, can de calculated as  $\lambda=r_2/r_1$, where $r_1$ and $r_2$ are the positions of the $g(r)$ subpeak maxima. The bond fraction, $\phi$ is determined as $\phi=n_1/(n_1+n_2)$ , where $n_1  = 4\pi\rho \int_0^{r_{m1}} {r^2 g(r)dr}$ and $n _2  = 4\pi\rho \int_{r_{m1}}^{r_{m2}} {r^2 g(r)dr}$ are respectively the effective numbers of short- and long-bonded particles in the first coordination shell. Here, $r_{m1}$ and $r_{m2}$ are locations of the first and the second $g(r)$ minima separating the subpeaks.

 In Fig.~\ref{fig:eff_param} we show typical liquid-phase RDFs for RSS at different values of parameter $\sigma$ which determines the bond length ratio $\lambda$. RSS is a useful model because it is very simple but reproduce all non-trivial effects in two-length-scale system~\cite{Fomin2008JCP,Gribova2009PRE,Ryltsev2013PRL,Ryltsev2015SoftMatt,Ryltsev2017SoftMatt,Fomin2019PhysChemLiq}. RDFs for RSS system presented in Fig.~\ref{fig:eff_param} reveal two important features typical for all two-length-scale systems. First, we see that relative height of sub-peaks and so the effective concentration $\phi$ can essentially change with density (pressure). Second, effective parameters are not always well-defined because RDF subpeaks corresponding to different bond lengths can overlap. Indeed, $\lambda$, $\phi$ are well defined in the case of $1.2 \gtrsim \lambda \lesssim 1.6$ then $g(r)$ subpeaks corresponding to short- and long-bonded particles are perfectly separated at arbitrary $\phi$ values (see Fig.~\ref{fig:eff_param}b).  But the situation is more complicated if $\lambda \lesssim 1.2$  or $\gtrsim 1.6$. In both cases there in no explicit splitting of the RDF peaks, only shoulders on first or second peaks are observed (see Fig.~\ref{fig:eff_param}a,c). In such cases, the effective parameters are ill-defined. However, we can estimate them by using the method of peak separation widely used in spectroscopy \cite{Butler1980PhotocemPotobiol,Arag2008JBrazChemSoc}. The method is based on using high order (2th and 4th) derivatives to separate overlapped peaks (see details in Ref.~\cite{Ryltsev2017SoftMatt}). Briefly, the maximum of $d^2 g(r)/dr^2$ allows estimating the distance $r_s$ corresponding to intersection of the subpeaks.  So the effective numbers of short- and long-bonded particles can be estimated as $n_1  = 4\pi\rho \int_0^{r_{m}} {r^2 g(r)dr}$ and $n _2  = 4\pi\rho \int_{r_{m}}^{r_{s}} {r^2 g(r)dr}$.

\section{Solid phase formation}

\subsection{Quasicrystals}

Earlier, we reported that the effective parameters described above can be successfully applied for predicting quasicrystal (QC) formation~\cite{Ryltsev2015SoftMatt,Ryltsev2017SoftMatt}. Here we describe briefly these results.

The method under consideration was firstly applied to decagonal quasicrystals~\cite{Ryltsev2015SoftMatt}. When studying the RSS system at $\sigma\in(1.3,1.4)$ we obtained that, in a certain density range, the system undergoes phase transition into either decagonal QC(DQC) or corresponding approximants. We guessed that the formation of such decagonal phases is a universal property of two-length-scales systems and so different systems would form the same phases at the parameters generating similar liquid structure (similar values of effective parameters). For RSS system, we obtain that the range of effective parameters corresponding to DQC formation is $\lambda\simeq(1.35-1.4)$ and $\phi\simeq (0.06-0.15)$.

To validate if these values are universal for other systems, we perform simulations with two alternative two-scale potentials: modified oscillating pair potential (OPPm)~\cite{Ryltsev2015SoftMatt} and Yoshida-Kamakura potential~\cite{Yoshida1976ProgTheorPhys}. Adjusting system parameters to obtain appropriate values of $(\lambda, \phi)$ in the liquid state, we cool the systems and observe the self-assembling of similar decagonal solid phases. That suggests the proposed criterion of decagonal structure formation is general and does not depend on any peculiarities of the system except the existence of two length-scales of the interaction. Of course, some more subtle features of decagonal phases such as the regions of QC stability or the structure of competing approximants may depend on particular system properties.

Then we showed that the criterion works well for the case of dodecagonal (12-fold) quasicrystals (DDQCs). To do so, we used four different two-length-scale potentials: Dzugutov potential \cite{Dzugutov1993PRL}, RSS, OPPm and the embedded-atom model (EAM) potential for aluminum proposed in \cite{Mishin1999PRB}. The values of effective parameters favoring dodecagonal order were determined from the system with Dzugutov potential for which the temperature-density domain of DDQC formation was known \cite{Dzugutov1993PRL}. Adjusting the states of RSS and OPP fluids to obtain the same values of effective parameters, we observed self-assembly of the same DDQC phases at cooling. The values of the parameters for the EAM model \cite{Mishin1999PRB} of liquid aluminum near the liquid-DDQC transition reported in \cite{Prokhoda2014arXiv} are also the same. This result suggests the common nature of both metallic and soft matter DDQCs arising from competition between length scales.

\subsection{Complex crystal phases}

The results presented above allow concluding that the method of effective parameters is a useful tool to predict the formation of quasicrystals of various symmetries in one-component two-length-scale systems. An important result is the universality of the method: the values of the effective parameters corresponding to the formation of a certain phase are close for different systems.

A natural question arises if the same universality exists for crystalline phases. The importance of the answer to this question is due to the fact that, in multiscale systems, a large number of various crystalline phases are usually observed. The concept of universality would help to classify the complex ''zoology'' of observed structures.

We begin the discussion of this issue with the case when effective parameters are well-defined from RDFs.

At this stage, we shall consider the formation of two crystalline phases: A3 hP2 (Mg) and cP4 (Li). The reasons for this choice are twofold: first, effective parameters are well-defined for these cases and second these phases were found in \cite{Englel2015Nature} for a one-component system with the OPP potential as competing with the icosahedral quasicrystal phase. Moreover, the cP4 phase has not been experimentally discovered so far. However, it has been theoretically predicted in \cite{Ma2008PRB} as one of the possible crystalline modifications of Li at high pressures.

\begin{figure}
  \centering
  \includegraphics[width=0.85\columnwidth]{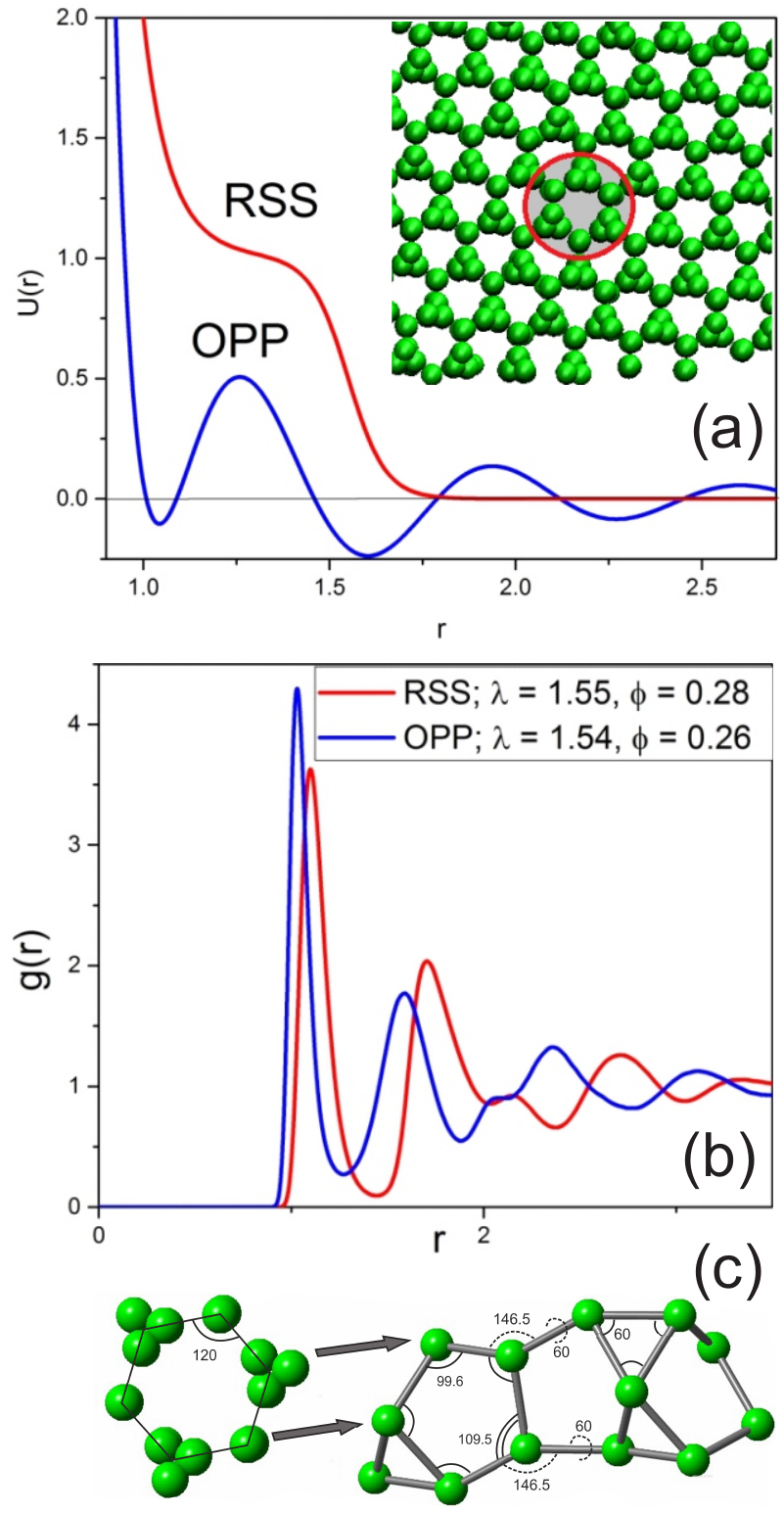}
 \caption{The universality of cP4(Li) structure formation. (a) Pair potentials for RSS and OPP generating liquids with close values of the effective parameters. Inset: the structure of cP4 phase formed by the cooling of RSS and OPP fluids in projection onto the (001) plane. (b) Radial distribution functions of RSS and OPP fluids and corresponding effective parameter values; (c) spatial structure of the hexagon highlighted by a circle in panel (a).}
  \label{fig:cP4_Li}
\end{figure}

\begin{figure}
  \centering
  \includegraphics[width=0.85\columnwidth]{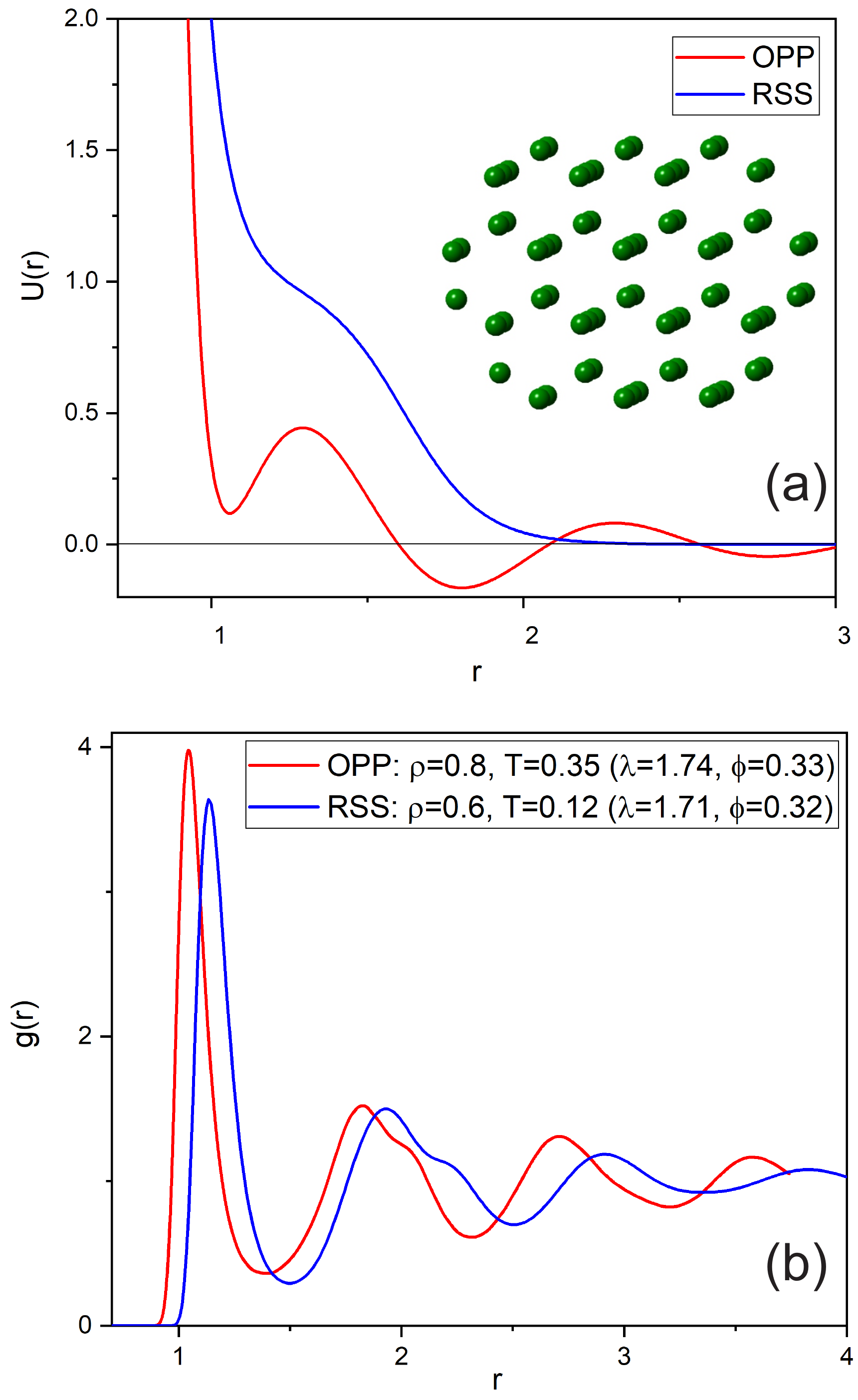}
 \caption{The universality of A3 (Mg) structure formation. (a) Pair potentials RSS and OPP generating liquids with close values of the effective parameters. (b) Radial distribution functions of RSS and OPP fluids and corresponding effective parameter values; Inset: A3 phase structure formed by the cooling of RSS and OPP fluids.}
  \label{fig:A3}
\end{figure}

Using the potential parameters given in \cite{Englel2015Nature}, we simulate the OPP system in the liquid state and estimate the effective parameters for the structures mentioned. The values obtained are equal to ($ \lambda\simeq 1.75, \phi\simeq 0.32) $ for hP2 phase and $(\lambda\simeq1.55, \phi\simeq0.25)$ for cP4 phase. A convenient system for validating the universal formation of these phases is RSS, which we studied earlier. For this system, it is easy to obtain the values of parameters that generate the given values of the effective parameters of the liquid. The parameter $\lambda $ (bond lengths ratio) is determined by the width $\sigma$ of the repulsive step of the potential. The parameter $\phi$ (concentration of short-bonded particles) is mainly determined by the density of the system. Several systems of various densities were simulated for fixed values of $\sigma = 1.55, 1.75$. As a result, liquids with the effective parameters close to those for the OPP system were obtained (see Fig.~\ref{fig:cP4_Li}, Fig.~\ref{fig:A3}. By cooling these liquids, the desired hP2 and cP4 structures were obtained (see Fig.~\ref{fig:cP4_Li} and Fig.~\ref{fig:A3}). This result shows that the method of effective parameters can be used to predict the formation of crystalline structures.

\subsection{Pressure-induced FCC-BCC transition}

Here we discuss the possibility to predict pressure-induced FCC-BCC transition from the structural characteristics of the fluid phase. This transition is a universal phenomenon observed in both two-lengths-scale systems and systems with soft repulsions at relatively low densities (pressures). The universality of this transition is caused by the fact that, at relatively low densities, the stable phase in any one-component system with pair isotropic potential is close-packed FCC (or HCP) crystalline lattice which has one characteristic length scale. However, in the presence of either the second length-scale of the potential or its softness, the increase of the pressure can cause the appearance of nonnegligible interactions with the second neighbors and thus stabilize the BCC structure which has two characteristic length scales. We should also notice that, in high-density limit, FCC phase is expected to be stable in any two-length-scale system with strong enough repulsion at short distances. Thus, two-length-scale systems often demonstrate FCC-BCC transition at low densities and BCC-FCC one at high densities~\cite{Fomin2008JCP,Prestipino2009SoftMatt}.

The effective parameters $(\lambda, \phi)$ for ideal BCC lattice can be easily obtained from its geometry: $(\lambda_{\rm bcc} = \sqrt{4/3}\approx1.15,\phi_{\rm bcc} = 0.5)$. So we expect that a liquid whose RDF generates the effective parameters close to these ideal values would crystallize at cooling into BCC phase. However, the application of this idea is difficult due to the fact that, at $\lambda \lesssim 1.2$, the RDF peaks corresponding to different bond lengths overlap and so the effective parameters are ill-defined (see Fig.~\ref{fig:eff_param}a). If the overlapping is not so substantial (pronounced shoulder on the RDF peak takes place), we can estimate th effective parameter $\lambda$  by using the method described briefly in the section~\ref{sec:eff_par}.

\begin{figure}
  \centering
  \includegraphics[width=0.95\columnwidth]{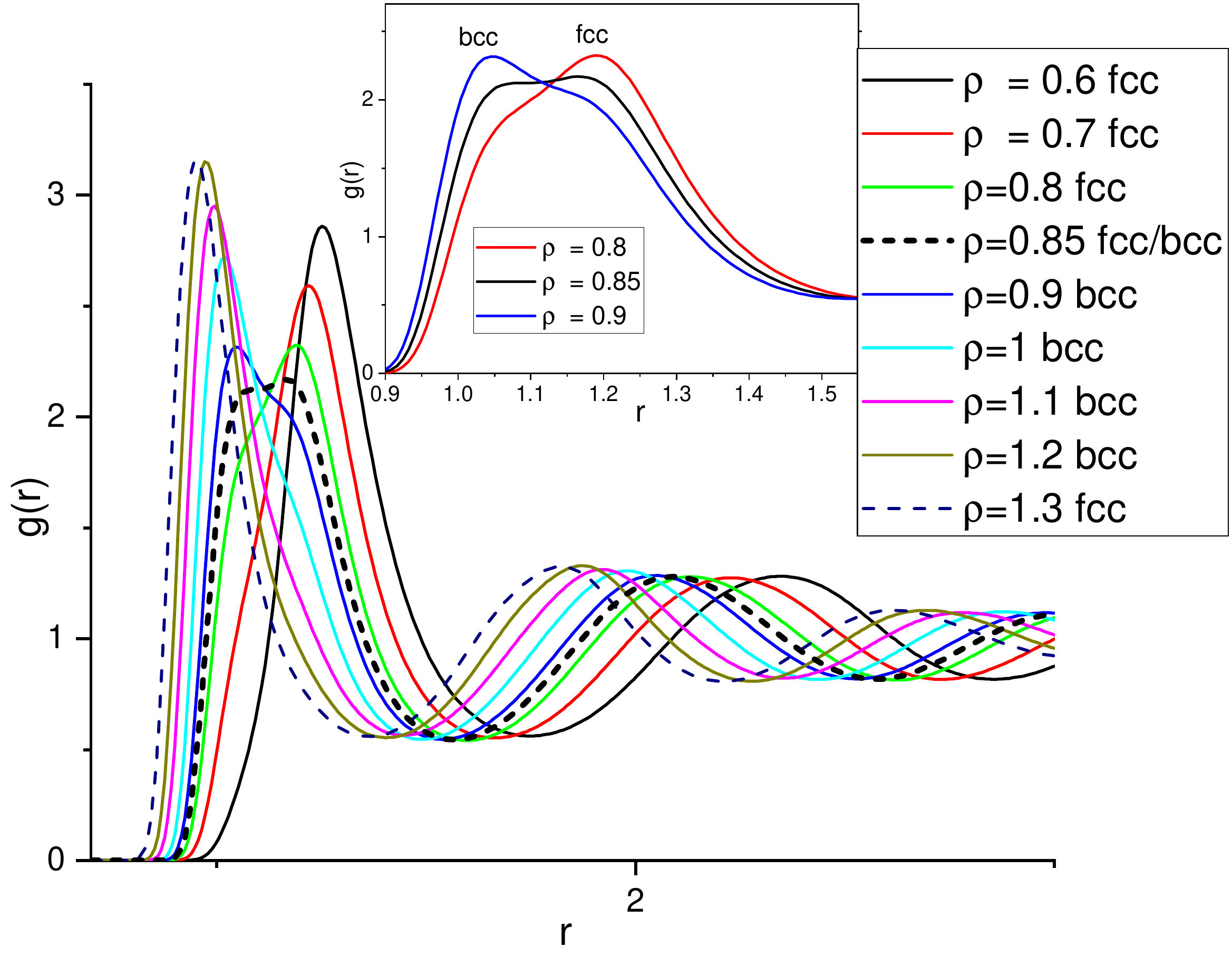} \\ \includegraphics[width=0.95\columnwidth]{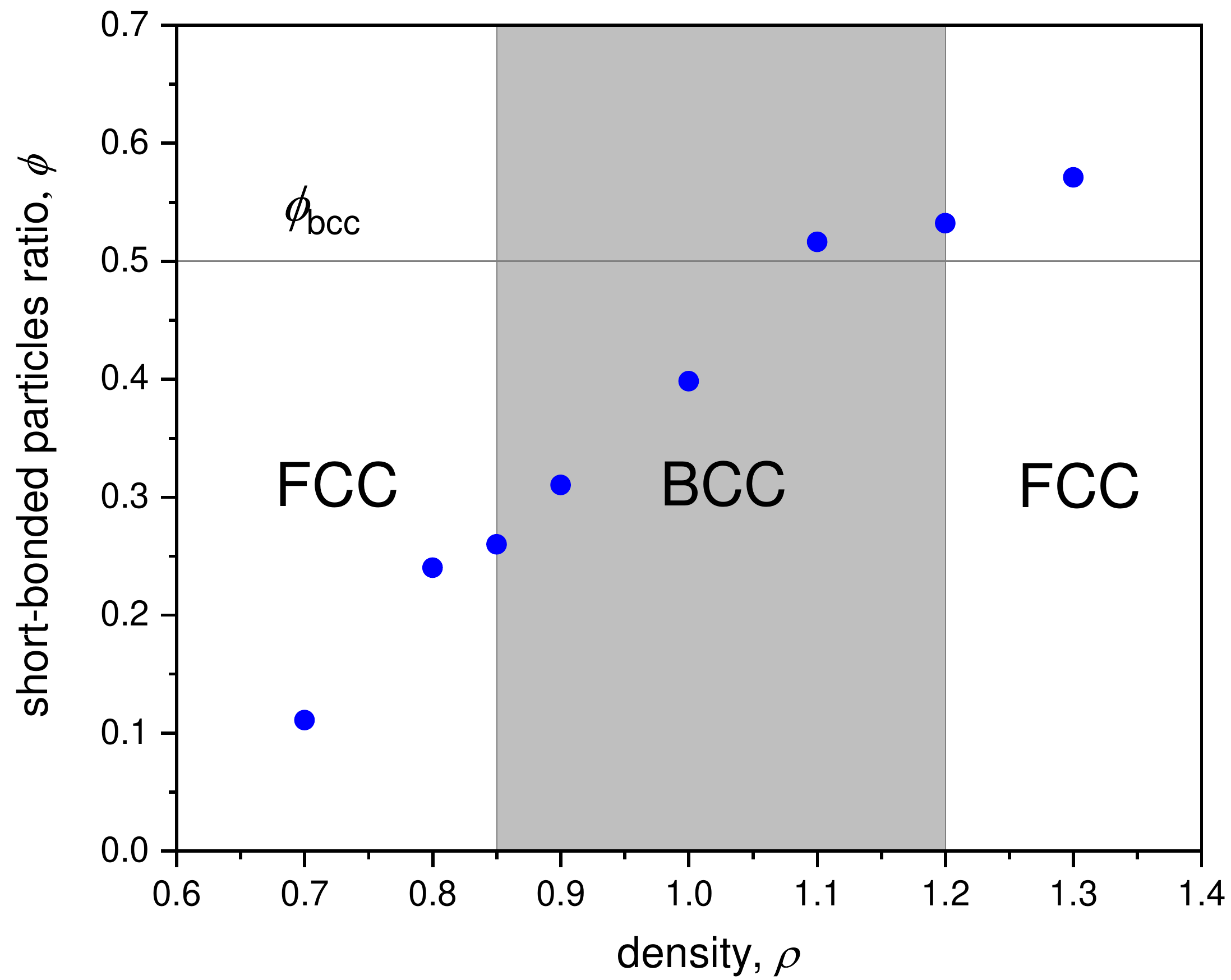}
 \caption{(a) Radial distribution functions for RSS system at $\sigma = 1.15$ at different densities. Each curve is calculated at the temperature corresponding to the melting of the solid phase formed at the freezing of the liquid. The inset shows RDFs in the vicinity of FCC-BCC transition; (b) Density dependence of short-bonded particles concentration $\phi$ for RSS melts at $\sigma=1.15$. We see that the density region corresponding to BCC phase formation is in the vicinity of $\phi_{\rm bcc}=0.5$.}
  \label{fig:rdf_fcc-bcc}
\end{figure}

To check the above idea, we perform simulations of freezing in RSS system at $\lambda = 1.15\approx\lambda_{\rm bcc}$. In this case, the pressure-temperature phase diagram has been calculated using the thermodynamic integration method~\cite{Fomin2008JCP}.  This diagram indeed reveals the existence of FCC-BCC-FCC transitions. In our simulations, this diagram was validated by direct observation of solid phase formation at the freezing of the fluid phase. In Fig.~\ref{fig:rdf_fcc-bcc} we show RDFs calculated at different densities and at corresponding melting temperatures. On the legend for each RDF curve in Fig.~\ref{fig:rdf_fcc-bcc} we specify the symmetry of the solid structures forming at the cooling of the liquid. We see a clear difference between liquid state RDFs calculated at the densities corresponding to the formation of FCC and BCC structures. Namely, the heights of the subpeaks corresponding to short-bonded and long-bonded particles change so that the first (doubled) peak transforms from left-shouldered to right-shouldered one as the density increases. That means the effective concentration of short-bonded particles increases with density.  This feature is the most pronounced in the density range near FCC-BCC transition which occurs at $\rho\approx0.85$ (see insert in Fig.~\ref{fig:rdf_fcc-bcc}). Note that, at this density, the liquid freezes into a mixture of FCC and BCC phases (remember that we use NVT ensemble). In this ''transition'' density range, we can estimate effective parameters because it is possible to detect the distance $r_s$ corresponding to the intersection of the subpeaks. The result of such estimation for $\phi$ is shown in the inset for  Fig.~\ref{fig:rdf_fcc-bcc}. We see that the density region corresponding to BCC phase formation is in the vicinity of $\phi_{\rm bcc}=0.5$.  Still, the bond length ratio $\lambda\approx\sigma=1.15$ is almost constant. Thus, the method of effective parameters can satisfactory predict FCC-BCC transition in the case when the distance $r_s$ corresponding to the intersection of the subpeaks can be determined (i.e. maximum on $d^2 g(r)/dr^2$ at $r=r_s$ takes place).

However, the method described above is not universally applied to any system demonstrating FCC-BCC transition because in certain cases different length scales can not be separated from the analysis of the RDF peaks. An illustrative example is the case of the systems interacting through one-length-scale ultrasoft repulsive potentials, such as Hertz and harmonic potentials~\cite{Fomin2018MolPhys,Levashov2017JCP,Levashov2019SoftMatt}. In such systems, fluid state RDFs in the vicinity of FCC-BCC transition do not demonstrate any shoulders; only widening of the first is observed~\cite{Levashov2019SoftMatt}.  In such cases, we can not use effective parameters and should apply some integral metrics to compare RDFs in different systems. The development of such metrics is the matter of separate work.

\section{Conclusions\label{sec:discuss}}

 Using molecular dynamics simulations we address general properties of freezing in one-component two-length-scale systems. Our findings allow concluding that the formation of solid phases during cooling a liquid is essentially determined by the radial correlation function (RDF) of the liquid. We show that different two-length-scale liquids having similar RDF freeze into the same solid phases. In certain cases, the similarity between RDFs can be expressed by the proximity of two dimensionless effective parameters: the ratio between effective bond lengths, $\lambda$, and the fraction of short-bonded particles $\phi$. We validate this idea in different two-length-scale systems by studying the formation of decagonal and dodecagonal quasicrystals, A3 an hP4 crystal phases as well as pressure-induced FCC-BCC transition. The methods developed allow predicting effectively the formation of solid phases in both numerical simulations and self-assembling experiments in soft matter systems with tunable interactions~\cite{Yakovlev2018IntSocOptPhot,Yakovlev2018JPhysConfSer}.

\section{Acknowledgments}
This work was supported by the Russian Science Foundation (grant 18-12-00438). The numerical calculations are carried out using computing resources of the federal collective usage center 'Complex for Simulation and Data Processing for Mega-science Facilities' at NRC 'Kurchatov Institute' (http://ckp.nrcki.ru/), supercomputers at Joint Supercomputer Center of Russian Academy of Sciences (http://www.jscc.ru) and 'Uran' supercomputer of IMM UB RAS (http://parallel.uran.ru).

%\bibliography{ourbib_gen}

%

\end{document}